\documentclass[aps,showpacs,preprintnumbers,amsmath, amssymb]{revtex4}

\usepackage{float}
\usepackage{graphics,epsfig}
\usepackage{graphicx}
\usepackage{dcolumn}
\usepackage{bm}

\begin{document}
\baselineskip=0.8 cm

\title{{\bf Scalarization of horizonless reflecting stars: neutral scalar fields non-minimally coupled to Maxwell fields}}
\author{Yan Peng$^{1}$\footnote{yanpengphy@163.com}}
\affiliation{\\$^{1}$ School of Mathematical Sciences, Qufu Normal University, Qufu, Shandong 273165, China}

\vspace*{0.2cm}
\begin{abstract}
\baselineskip=0.6 cm
\begin{center}
{\bf Abstract}
\end{center}

We analyze condensation behaviors of neutral
scalar fields outside horizonless reflecting stars
in the Einstein-Maxwell-scalar gravity.
It was known that minimally coupled neutral scalar fields
cannot exist outside horizonless reflecting stars.
In this work, we consider non-minimal couplings
between scalar fields and Maxwell fields, which is included to aim to
trigger formations of scalar hairs. We analytically demonstrate that
there is no hair theorem for small coupling parameters below a bound.
For large coupling parameters above the bound, we numerically
obtain regular scalar hairy configurations supported by horizonless
reflecting stars.

\end{abstract}

\maketitle
\newpage
\vspace*{0.2cm}

\section{Introduction}

The recent observation of gravitational waves
may provide a way to test the nature of
astrophysical black holes \cite{BP1,BP2,BP3}.
In classical general relativity, one famous
property of black holes is no hair theorem, which states
that a nontrivial static scalar field
cannot exist in the exterior
region of asymptotically flat black holes,
see references \cite{Bekenstein}-\cite{sn3} and
reviews \cite{Bekenstein-1,CAR}.
However, some candidate quantum-gravity models suggested that,
due to quantum effects \cite{UH1,UH2,UH3,UH4,UH5},
the classical absorbing horizon should be replaced by
a reflecting surface \cite{RS1,RS2,RS3,RS4,RS5,RS6,RS7}.
Interestingly, no hair behaviors also appear
for such horizonless reflecting stars \cite{Hod-1}-\cite{BKM}.
In particular, even for static scalar fields non-minimally coupled to the Ricci curvature,
no hair theorem still holds in backgrounds of black holes and horizonless
reflecting stars \cite{nm1,nm2,nm3,Hod-5,Hod-6,nm4}.

Intriguingly, for static scalar fields non-minimally coupled to the Gauss-Bonnet invariant,
scalar field hairs can exist outside
asymptotically flat black holes \cite{SGB1,SGB2,SGB3,SGB4}.
 Spinning hairy black holes were also numerically obtained in the scalar-Gauss-Bonnet theory \cite{SGB5}.
In addition, analytical formula of the scalar-Gauss-Bonnet coupling parameter was explored in \cite{SGB6}.
These scalarization models are constructed by introducing an additional term $f(\psi)R_{GB}^2$,
where $f(\psi)$ is a function of the scalar field $\psi$
and $R_{GB}^2$ is the Gauss-Bonnet invariant.
In the scalar-Gauss-Bonnet gravity, under scalar perturbations,
the bald black hole is thermodynamically unstable and it may
evolve into a hairy black hole \cite{SGB2,SGB3}.
This intriguing mechanism of hair formations is usually
called spontaneous scalarization,
which was found long ago for neutron stars in the context
of scalar-tensor theories \cite{SGBa}.
At present, lots of spontaneous scalarization models
were constructed in the background of black holes
\cite{SGB8,SGB9,SGB10,SGB11,SGB12,SGB13,SGB14,SGB15,SGB16}.

As mentioned above, some candidate quantum-gravity models suggested that
quantum effects may prevent the formation of horizons and a reflecting
wall may lay above the would-be horizon position \cite{UH1}-\cite{RS3}.
Interestingly, it was found that neutral scalar field hairs
cannot exist outside such horizonless reflecting stars (even the
star is charged )\cite{Hod-1}. When considering scalar-Gauss-Bonnet couplings,
we showed that the coupling can lead to the formation of
neutral scalar field hairs in the background of
horizonless reflecting stars \cite{ss1}.
In fact, black hole spontaneous scalarization is a very
universal property, which also can  be induced by another type of non-minimal
couplings between scalar fields and Maxwell fields \cite{charge1,charge2,charge3,charge4,charge5,charge6}.
As a further step, it is very interesting to
examine whether scalar-Maxwell couplings can trigger
condensations of neutral scalar fields outside
horizonless reflecting stars.

This work is organized as follows. We start by introducing
a model with a neutral scalar field coupled to the Maxwell field
in the charged horizonless reflecting star spacetime.
For small coupling parameters, the neutral scalar field
cannot exist. In contrast, for large coupling parameters, we
get numerical solutions of scalar hairy horizonless
reflecting stars. Main conclusions
are presented in the last section.

\section{Investigations on the coupling parameter between scalar fields and Maxwell fields}

We take the Lagrange density with scalar fields non-minimally coupled
to Maxwell fields in the asymptotically flat background.
It is defined by the following expression \cite{charge1,charge2,charge3}
\begin{eqnarray}\label{lagrange-1}
\mathcal{L}=R-\nabla^{\nu}\nabla_{\nu}\Psi-\mu^{2}\psi^{2}+f(\Psi)\mathcal{I}.
\end{eqnarray}
Here R is the scalar curvature.
$\Psi$ is the static neutral scalar field with mass $\mu$.
$f(\Psi)$ is a function coupled
to $\mathcal{I}=F_{\rho\sigma}F^{\rho\sigma}$.
In the linearized regime, there is $\mathcal{I}=-\frac{Q^2}{r^4}$ and
the general coupling function can be expressed as $f(\Psi)=1-\alpha\Psi^2$,
where $\alpha$ is the model parameter describing coupling strength \cite{charge1,charge2,charge3}.
In the limit of $\alpha\rightarrow 0$, it returns to the usual
Einstein-Maxwell-scalar gravity.

The scalar field differential equation is
\begin{eqnarray}\label{lagrange-1}
\nabla^{\nu}\nabla_{\nu}\Psi-\mu^2\Psi+\frac{f'_{\Psi}\mathcal{I}}{2}=0.
\end{eqnarray}

The charged static spherically symmetric background is
\begin{eqnarray}\label{AdSBH}
ds^{2}&=&-N(r)dt^{2}+\frac{dr^{2}}{N(r)}+r^{2}(d\theta^{2}+sin^{2}\theta d\phi^{2}).
\end{eqnarray}
In the weak-field limit, the metric function $N(r)$ is
\begin{eqnarray}\label{AdSBH}
N(r)=1-\frac{2M}{r}+\frac{Q^2}{r^2}
\end{eqnarray}
with M and Q representing the star mass and star charge respectively.
We point out that this background metric is valid on the condition $\alpha\psi^2\ll 1$

We take the scalar field decomposition
\begin{eqnarray}\label{BHg}
\Psi(r,\theta,\phi)=\sum_{lm}e^{im\phi}S_{lm}(\theta)R_{lm}(r).
\end{eqnarray}

For simplicity, we label $R_{lm}(r)$ as $\psi(r)$.
With relations (2), (3), (5) and $\mathcal{I}=-\frac{Q^2}{r^4}$, we derive the ordinary differential equation \cite{Hod-4,hin1,hin2}
\begin{eqnarray}\label{BHg}
\psi''+(\frac{2}{r}+\frac{N'}{N})\psi'+(\frac{\alpha Q^2}{r^4N}-\frac{l(l+1)}{r^2N}-\frac{\mu^2}{N})\psi=0.
\end{eqnarray}
Here $l$ is the spherical harmonic index and $l(l+1)$ is the characteristic
eigenvalue of the angular scalar eigenfunction $S_{lm}(\theta)$.

We label $r_{s}$ as the radial coordinate of the star surface.
Since we focus on the compact star without a horizon,
the star surface is outside the gravitational
radius, which can be expressed as $r_{s}> M+\sqrt{M^2-Q^2}$.
At the star surface, we take scalar reflecting surface boundary conditions
$\psi(r_{s})=0$. In the far region, the physical massive static scalar fields
asymptotically behave as $\psi(r\rightarrow \infty)\sim \frac{1}{r}e^{-\mu r}$.
So the scalar field satisfies bound-state conditions
\begin{eqnarray}\label{InfBH}
&&\psi(r_{s})=0,~~~~~~~~~\psi(\infty)=0.
\end{eqnarray}

According to boundary conditions (7), one concludes that the
function $\psi(r)$ must possess (at least) one extremum point $r=r_{peak}$
between the star surface $r=r_{s}$ and spatial infinity.
It can be a positive maximum extremum point or a negative minimum extremum point.
With the symmetry $\psi\rightarrow -\psi$ of equation (6), without loss of generality,
we can only study the case of positive maximum extremum points.
Then the scalar field around the extremum point is
characterized by \cite{Hod-1}
\begin{eqnarray}\label{InfBH}
\psi(r_{peak})>0,~~ \psi'(r_{peak})=0,~~ \psi''(r_{peak})\leqslant0.
\end{eqnarray}

(6) and (8) yield the relation
\begin{eqnarray}\label{BHg}
\frac{\alpha Q^2}{r^4N}-\frac{l(l+1)}{r^2N}-\frac{\mu^2}{N}\geqslant 0~~~for~~~r=r_{peak}.
\end{eqnarray}

Since we concentrate on horizonless stars,
the extremum point is outside the gravitational radius satisfying
\begin{eqnarray}\label{BHg}
N(r_{peak})=1-\frac{2M}{r_{peak}}+\frac{Q^2}{r_{peak}^2}>0.
\end{eqnarray}

With (9) and (10), we get the relation
\begin{eqnarray}\label{BHg}
\frac{\alpha Q^2}{r_{peak}^4}-\frac{l(l+1)}{r_{peak}^2}-\mu^2\geqslant 0.
\end{eqnarray}

According to the inequality (11), we deduce a bound on the coupling parameter
\begin{eqnarray}\label{BHg}
\alpha\geqslant \frac{\mu^2r_{peak}^4+l(l+1)r_{peak}^2}{Q^2}\geqslant \frac{\mu^2r_{s}^4+l(l+1)r_{s}^2}{Q^2}
>\frac{\mu^2(M+\sqrt{M^2-Q^2})^4+l(l+1)(M+\sqrt{M^2-Q^2})^2}{Q^2}.
\end{eqnarray}

If compact reflecting stars are surrounded with static neutral scalar hairs,
the parameter $\alpha$ should be above the bound (12).
In other words, we obtain a no hair theorem
for small coupling parameters
\begin{eqnarray}\label{BHg}
\alpha\leqslant \frac{\mu^2(M+\sqrt{M^2-Q^2})^4+l(l+1)(M+\sqrt{M^2-Q^2})^2}{Q^2}.
\end{eqnarray}
It implies that neutral massive static exterior scalar fields usually cannot
exist in cases of large field mass, large  spherical harmonic index,
large star mass or small star charge.
In particular, for $\alpha=0$, the no hair condition (13) always
holds, which means that charged stars cannot support
minimally coupled neutral scalar hairs.
In the following, we numerically show that scalar hairs
can be induced by large non-minimal coupling parameters satisfying (12).

\section{Neutral scalar field hairs non-minimally coupled to Maxwell fields}

We numerically solve the equation (6)
together with boundary conditions (7). Besides parameters $r_{s}$, $M$, $Q$, $l$
and $\alpha$, we also need initial values of $\psi(r_{s})$ and $\psi'(r_{s})$
to integrate the equation. The reflecting condition of (7) gives the value $\psi(r_{s})=0$.
According to the symmetry $\psi\rightarrow k \psi$ of equation (6),
we firstly set $\psi'(r_{s})=1$ without loss of generality.
Since the equation (6) also satisfies
the symmetry $r\rightarrow \gamma r,~ \mu\rightarrow \mu/\gamma,~ M\rightarrow \gamma M,~ Q\rightarrow \gamma Q$,
we use dimensionless parameters $\mu r_{s}$, $\mu M$, $\mu Q$, $l$
and $\alpha$ to describe the system. For given values of $\mu r_{s}$, $\mu M$, $\mu Q$ and $l$,
using standard shooting methods, we search for the proper $\alpha$ with the
vanishing condition $\psi(\infty)=0$.
The equation of motion of the scalar field is linear with respect to $\psi$ and the solution is scale free.
After getting numerical solutions, we modify the boundary condition $\psi'(r_{s})=1$ so that
$\alpha\psi^2\ll 1$ holds. In this work, we take very
small $\alpha\psi^2$ satisfying $\alpha\psi^2<10^{-7}$.

In the case of $\mu r_{s}=2.7$, $\mu M=1.5$, $\mu Q=1.0$ and $l=0$, we choose
various $\alpha$ to try to get the physical
solution with $\psi(\infty)=0$.
As shown by red curves in Fig. 1, if we choose $\alpha=322$,
the solution diverges quickly to be $\infty$.
For green curves in Fig. 1, if we choose a little larger value $\alpha=323$,
then the solution decreases to be $-\infty$ in the larger r region.
It turns out that $\alpha=322$ and $\alpha=323$ are not related to the physical
scalar field solution with decaying behaviors at infinity.

\begin{figure}[h]
\includegraphics[width=230pt]{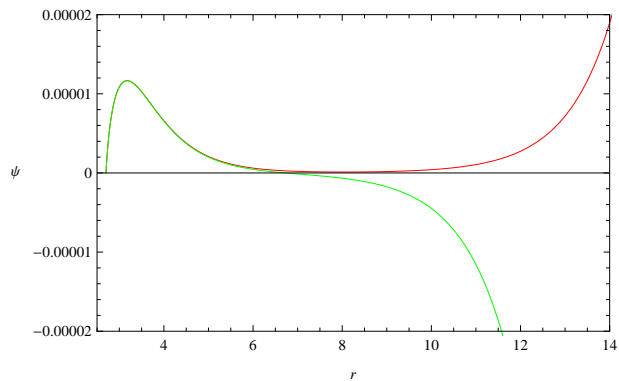}\
\caption{\label{EEntropySoliton} (Color online) We show behaviors of
$\psi(r)$ in cases of $\mu r_{s}=2.7$, $\mu M=1.5$, $\mu Q=1.0$, $l=0$
and different values of $\alpha$. The red line corresponds to $\alpha=322$ and the green line is with
$\alpha=323$. }
\end{figure}

In fact, general mathematical solutions of equation (6) behave as
$\psi\thickapprox A\cdot\frac{1}{r}e^{-\mu r}+B\cdot\frac{1}{r}e^{\mu r}$ with $r\rightarrow \infty$.
The red line of Fig. 1 corresponds to $B>0$ and the green line of Fig. 1
represents the case of $B<0$. As the value B should change continuously
with $\alpha$, indicating the existence of a critical
$\alpha$ corresponding to $B=0$. For this critical
$\alpha$, physical scalar fields asymptotically decay as
$\psi\varpropto \frac{1}{r}e^{-\mu r}$ at infinity.
With $\mu r_{s}=2.7$, $\mu M=1.5$, $\mu Q=1.0$ and $l=0$, we numerically
obtain a discrete value $\alpha\thickapprox322.083016$,
which corresponds to the solution satisfying $\psi(\infty)=0$.
We plot the physical solution with blue curves in Fig. 2,
which asymptotically approaches zero in the far region.
For higher modes $l\geq1$, we showed physical solutions with discrete $\alpha$ in Fig. 3.
Similarly, in other cases of black holes,
scalar hairy configurations are also characterized by
discrete coupling parameters in the linearized regime \cite{SGB3}.

\begin{figure}[h]
\includegraphics[width=230pt]{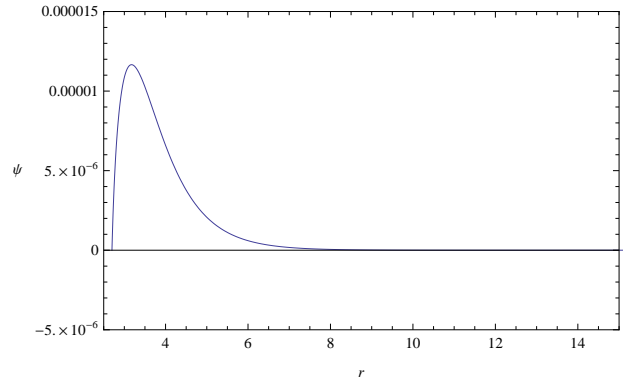}\
\caption{\label{EEntropySoliton} (Color online) We show the
function $\psi(r)$ in the case of $\mu r_{s}=2.7$, $\mu M=1.5$, $\mu Q=1.0$, $l=0$
and $\alpha=322.083016$. }
\end{figure}

\begin{figure}[h]
\includegraphics[width=180pt]{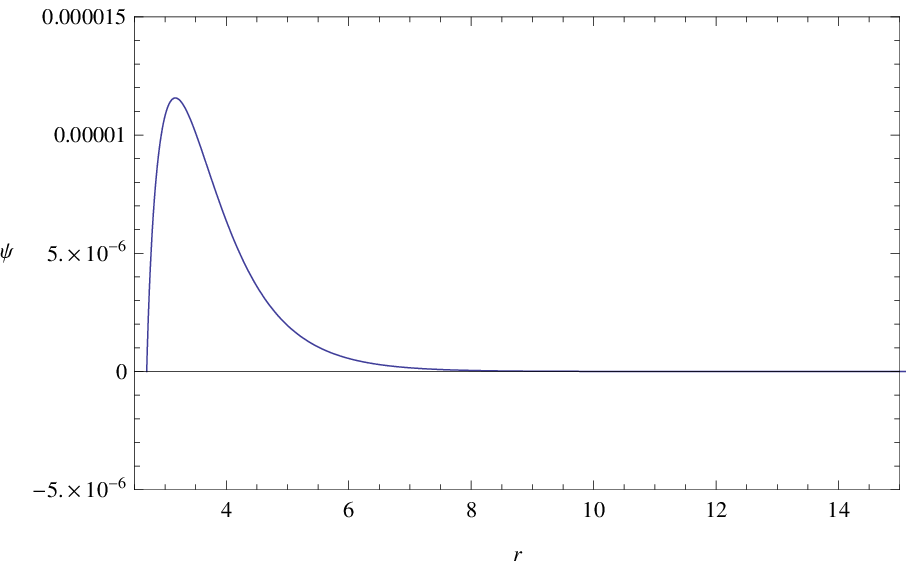}\
\includegraphics[width=180pt]{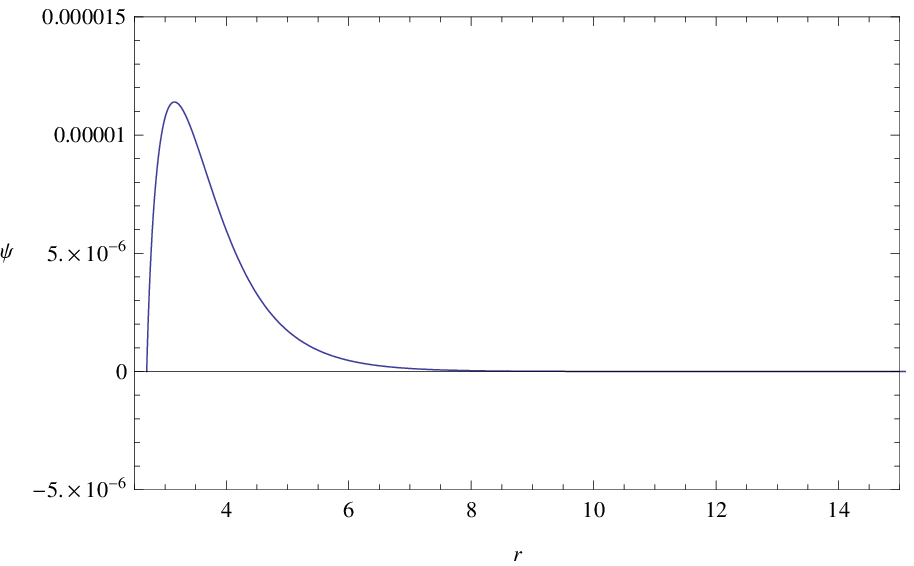}\
\caption{\label{EEntropySoliton} (Color online) We Plot the
function $\psi(r)$ in the case of $\mu r_{s}=2.7$, $\mu M=1.5$ and $\mu Q=1.0$.
The left panel corresponds to $l=1$ and $\alpha=347.375787$.
The right panel is with $l=2$ and $\alpha=397.388732$. The two physical solutions
with different $l$ behave very similarly to each other.}.
\end{figure}

Now we study how parameters $\mu r_{s}$, $\mu M$, $\mu Q$ and $l$ can affect
the discrete coupling parameter $\alpha$, which corresponds
to the decaying scalar field.
In Table I, for $\mu M=1.5$, $\mu Q=1.0$ and fixed $l$, we show effects of
$\mu r_{s}$ on discrete $\alpha$. It can be seen that
a larger radius $\mu r_{s}$ corresponds to a larger
discrete $\alpha$. With $\mu r_{s}=2.7$, $\mu Q=1$ and fixed $l$, according to data
in Table II, the discrete $\alpha$ decreases as we choose a larger
star mass $\mu M$. In Table III, we see that
the discrete $\alpha$ decreases as a function of the
star charge $\mu Q$.
Results in Table III also implies that $\alpha$ becomes smaller
when we choose a small $\overline{Q}=\frac{Q}{M}=\frac{\mu Q}{\mu M}$,
which is qualitatively the same as cases of black holes expressed by
analytical formula (17) in \cite{charge3}.
From data in Tables I, II and III, we see that larger
spherical harmonic index $l$ leads to a larger discrete coupling
parameter $\alpha$.

\renewcommand\arraystretch{2.0}
\begin{table} [h]
\centering
\caption{The parameter $\alpha(l)$ with $\mu M=1.5$, $\mu Q=1.0$ and various $\mu r_{s}$}
\label{address}
\begin{tabular}{|>{}c|>{}c|>{}c|>{}c|>{}c|>{}c|}
\hline
$~\mu r_{s}~$ & ~2.62~& ~2.66~& ~2.70~& ~2.74~& ~2.78\\
\hline
$~\alpha(l=0)~$ & ~258.102979~& ~298.304665~& ~322.083016~& ~343.522473~& ~364.215671\\
\hline
$~\alpha(l=1)~$ & ~280.963138~& ~322.681432~& ~347.375787~& ~369.638470~& ~391.118639\\
\hline
$~\alpha(l=2)~$ & ~326.150845~& ~370.872697~& ~397.388732~& ~421.291276~& ~444.340681\\
\hline
\end{tabular}
\end{table}

\renewcommand\arraystretch{2.0}
\begin{table} [h]
\centering
\caption{The parameter $\alpha(l)$ with $\mu r_{s}=2.7$, $\mu Q=1.0$ and various $\mu M$}
\label{address}
\begin{tabular}{|>{}c|>{}c|>{}c|>{}c|>{}c|>{}c|}
\hline
$~\mu M~$ &~1.48~& ~1.49~& ~1.50~& ~1.51~& ~1.52\\
\hline
$~\alpha(l=0)~$ & ~329.865858~ & ~326.322158~& ~322.083016~& ~316.754944~& ~309.433789\\
\hline
$~\alpha(l=1)~$ & ~355.357693~& ~351.725147~& ~347.375787~& ~341.903046~& ~334.372292\\
\hline
$~\alpha(l=2)~$ & ~405.760364~& ~401.953940~& ~397.388732~& ~391.632426~& ~383.690319\\
\hline
\end{tabular}
\end{table}

\renewcommand\arraystretch{2.0}
\begin{table} [h]
\centering
\caption{The parameter $\alpha(l)$ with $\mu r_{s}=2.7$, $\mu M=1.5$ and various $\mu Q$}
\label{address}
\begin{tabular}{|>{}c|>{}c|>{}c|>{}c|>{}c|>{}c|}
\hline
$~\mu Q~$ & ~0.92~& ~0.96~& ~1.00~& ~1.04~& ~1.08\\
\hline
$~\alpha(l=0)~$ & ~351.506957~& ~339.345528~& ~322.083016~& ~304.458791~& ~287.540951\\
\hline
$~\alpha(l=1)~$ & ~380.541786~& ~366.508043~& ~347.375787~& ~328.019191~& ~309.520394\\
\hline
$~\alpha(l=2)~$ & ~437.965671~& ~420.222598~& ~397.388732~& ~374.603527~& ~352.976403\\
\hline
\end{tabular}
\end{table}

\section{Conclusions}

We investigated formations of neutral scalar field hairs outside
asymptotically flat spherical horizonless reflecting stars.
We considered scalar fields non-minimally coupled to
Maxwell fields. We showed that the coupling parameter plays
an important role in scalar condensations.
We analytically got a bound on the coupling parameter
expressed in the form
$\alpha\leqslant \frac{\mu^2(M+\sqrt{M^2-Q^2})^4+l(l+1)(M+\sqrt{M^2-Q^2})^2}{Q^2}$,
where $\alpha$ is the coupling parameter, $\mu$ is the scalar
field mass, M is the star mass, Q is the star charge
and $l$ is the spherical harmonic index.
For $\alpha$ below this bound, no neutral scalar hair theorem holds.
In contrast, for large $\alpha$ above this bound,
with shooting methods, we obtained regular scalar hairy
configurations with a horizonless reflecting star in the center.
We also examined effects of star radii, star mass, star charge and spherical harmonic index
on condensations of neutral scalar fields.

\begin{acknowledgments}

This work was supported by the Shandong Provincial Natural Science Foundation of China under Grant
No. ZR2018QA008. This work was also supported by a grant from Qufu Normal University of China under Grant
No. xkjjc201906.

\end{acknowledgments}

\end{document}